# What defines stationarity in space plasmas

G. Livadiotis & D.J. McComas

*Department of Astrophysical Sciences, Princeton University, Princeton, NJ 08544, USA*

## Abstract

Starting from the concept of entropy defect in thermodynamics, we construct the entropy formulation of space plasmas, and then use it to develop a measure of their stationarity. In particular, we show that statistics of this entropy results in two findings that improve our understanding of stationary and nonstationary systems: (i) variations of the Boltzmann-Gibbs (BG) entropy do not exceed twice the value of the thermodynamic kappa, the parameter that provides a measure of the entropy defect in both stationary and nonstationary states, while becomes the shape parameter that labels the kappa distributions in stationary states; and (ii) the ratio of the deviation of the BG entropy with kappa scales with the kappa deviation via a power-law, while the respective exponent provides a stationarity deviation index (*SDI*), which measures the natural tendency of the system to depart from stationarity. We confirm the validity of these findings in three different heliospheric plasma datasets observed from three missions: (1) A solar energetic particle event, recorded by the Integrated Science Investigation of the Sun instrument onboard Parker Solar Probe; (2) Near Earth solar wind protons recorded by the Solar Wind Experiment instrument onboard WIND; and (3) Plasma protons in the heliosheath, source of energetic neutral atoms recorded by IBEX. The full strength and capability of the entropic deviation ratio and *SDI* can now be used by the space physics community for analyzing and characterizing the stationarity of space plasmas, as well as other researchers for analyzing any other correlated systems.

## 1. Introduction

This paper develops a measure that characterizes the stationarity of systems described by kappa distributions, such as space plasmas (e.g., Livadiotis & McComas 2009; 2010a,b; 2013a). While these systems reside in stationary states out of the classical thermodynamic equilibrium, they are still consistent with thermodynamics (e.g., Abe 2001; Livadiotis 2018a; Livadiotis & McComas 2022). However, as they move or flow the plasmas evolve, continuously subject to variations of their kappa distributions and associated thermodynamics, revealing a quasi-stationary behavior, that can also develop nonstationary features. Currently, there is no quantification of this natural tendency to leave stationarity and the purpose of this paper is to develop such a measure. For this, we revisit the kappa distributions and their thermodynamic origin, the concept of entropy defect (e.g., Livadiotis & McComas 2023a;b).

Space plasmas, from the solar wind and the planetary magnetospheres to the outer heliosphere and even interstellar and galactic plasmas beyond, are particle systems with long-range interactions whose energy and velocities are described by kappa distributions; (observations began in the 1960s; see the original studies by Binsack 1966; Olbert 1968; Vasyliūnas 1968; for the plethora of modern applications, see: the books: Livadiotis 2017 and Yoon 2019; and the reviews by Livadiotis & McComas 2009; 2013a; Pierrard & Lazar 2010; Livadiotis 2015a; Tsallis 2023). Understanding these kappa distributed plasmas is driving a whole new framework of thermodynamics (Abe 2001; Livadiotis 2018a; Livadiotis & McComas 2010a; 2021; 2022; 2023a).

Kappa distributions comprise the most generalized distribution functions of particle velocities for systems residing in stationary states, and is called generalized thermal equilibrium (Livadiotis & McComas 2009; 2013a; Livadiotis 2018a;b). This is in contrast to classical thermal equilibrium, which is a limiting version of equilibrium where there are no long-range interactions, $\kappa \rightarrow \infty$, and the Maxwell (1860) - Boltzmann (1866) (MB) distributions characterize the particles (Livadiotis & McComas 2013a;b). Kappa distributions have a thermodynamic origin, and are also connected to nonextensive statistical mechanics (e.g., see: Treumann 1997; Milovanov & Zelenyi 2000; Leubner 2002; Livadiotis & McComas 2009; 2010a; Livadiotis 2014). The kappa distributions have been used to determine the temperature and other key thermodynamic parameters of particle populations in space plasmas throughout the heliosphere (e.g., see: Livadiotis 2015a,b,c; 2017; Livadiotis et al. 2022; 2024a). (With regard to solar wind particles, some examples are the following: Maksimovic et al. 1997; Pierrard et al. 1999; Mann et al. 2002; Marsch 2006; Zouganelis 2008; Livadiotis & McComas 2010a; Yoon et al. 2012; Nicolaou & Livadiotis 2016; Pavlos et al. 2016; Livadiotis 2018c,d; Livadiotis et al. 2018; Nicolaou et al. 2019; Wilson et al. 2019; Silveira et al. 2021; Benetti et al. 2023.)

There are many detailed plasma physics mechanisms that can generate kappa distributions in space plasmas. These include superstatistics (e.g., Beck & Cohen 2003; Schwadron et al. 2010; Livadiotis et al. 2016; Gravanis et al. 2020; Davis et al. 2023; Ourabah 2024), the effect of shock waves (e.g., Zank et al. 2006), turbulence (e.g., Bian et al. 2014; Yoon 2014), acceleration mechanisms (e.g., Fisk & Gloeckler 2014), clustering (Peterson et al. 2013; Livadiotis et al. 2018); turbulent heating and polytropic processes (e.g., Livadiotis 2019; Nicolaou & Livadiotis 2019), entropic transfer by pickup ions (Livadiotis & McComas 2011a,b; 2023c; Livadiotis et al. 2024b).

Only particle distributions consistent with thermodynamics can maintain their form through the system's evolution over time. Systems residing in stationary states were shown to be consistently described by thermodynamics and only kappa distributions describe these stationary systems (e.g., Livadiotis 2018a; Livadiotis & McComas 2022; 2023a). Therefore, while each of the aforementioned mechanisms can generate a kappa distribution, it is the consistency of these developed distributions with thermodynamics that allows their extended existence in space plasmas. The thermodynamic property that produces kappa distributions instead of the classical case of MB distributions is the existence of particle correlations. In fact, the primary parameter, kappa, that labels and shapes these distributions is a physical measure of the amount of correlation, with the inverse kappa, $1/\kappa$, exactly equal to the correlation (normalized covariance) $\rho_\varepsilon$ of the kinetic energies of two particles, per half degrees of freedom (dof), $D$, i.e., $1/\kappa = \rho_\varepsilon /(\frac{1}{2}D)$ (Abe 1999; Asgarani & Mirza 2007; Livadiotis & McComas 2011a; 2021; 2023a; Livadiotis 2015a; Nelson et al. 2017; Livadiotis et al. 2021). Interestingly, this is parallel to the other thermodynamic parameter, the temperature, which is equal to the average kinetic energy $<\varepsilon>$ of any particle, per half dof, i.e., $k_B T = <\varepsilon>/(\frac{1}{2}D)$ (Livadiotis & McComas 2021; Livadiotis et al. 2021).

The classical statistical framework of Boltzmann (1866) - Gibbs (1902) (BG) has long been connected with thermodynamics. It requires that there are no correlations among particles, and as such, produces an additive formulation of entropy, the BG entropy. Once maximized within the framework of canonical ensemble, the BG entropy leads to the formalism of MB distributions. However, if the particles are not fully independent, the additivity has to be modified to account for correlations among the particles and the distribution is not MB; this is exactly how space plasmas (and many other systems) are described.

Space plasmas are examples of systems with correlations among their particles. For these, a simple summation does not characterize their entropy. Instead, Livadiotis & McComas (2023a;b;d) demonstrated that there is a certain way and mathematical formalism that describes such additions, including their correlations, which is consistent with thermodynamics. This provides a new addition rule for entropies that generalizes the classical simple summation. The generalized addition rule can be physically understood through the concept of entropy defect.

The entropy defect quantifies the change in entropy caused by the order induced in a system through the additional correlations among its constituents (e.g., particles) when these are ensembled together to form the combined system. It measures the missing entropy between the system and the sum of the individual entropies of its parts. The entropy defect is analogous to the mass defect, which quantifies the missing mass associated with assembling subatomic particles; similarly, the entropy defect quantifies the missing entropy associated with assembling particles with correlations, such as space plasmas (Figure 1).

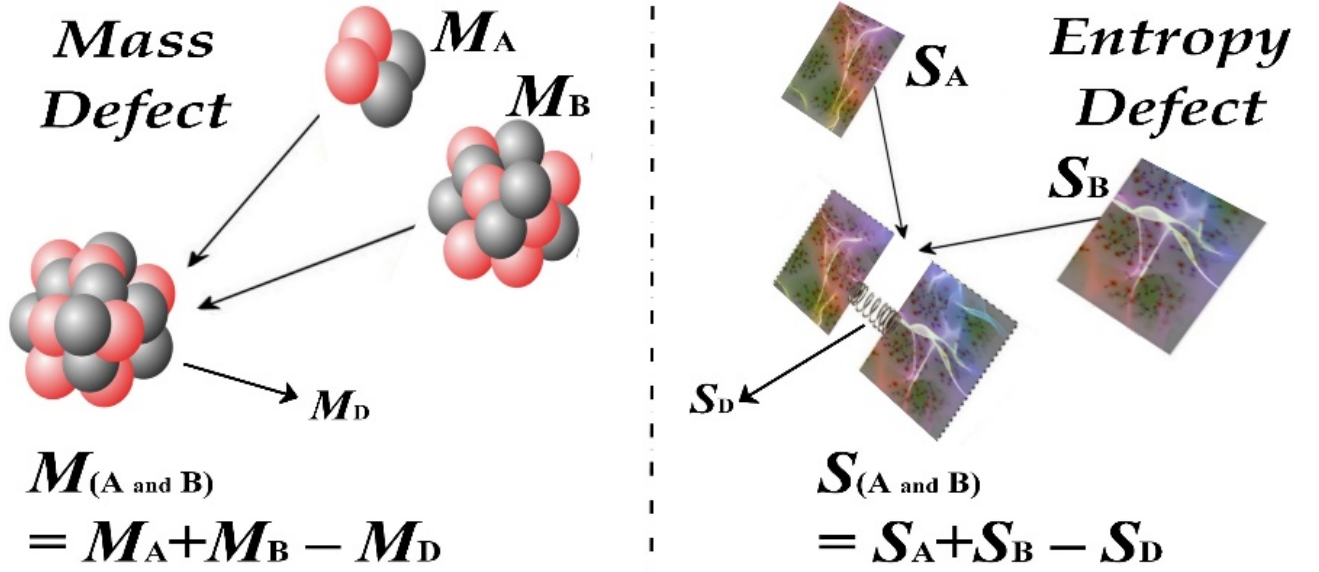

**Figure 1.** Schematic diagram comparing mass and entropy defects. Analogous to the mass defect ($M_\text{D}$) that quantifies the missing mass (energy) associated with assembling subatomic particles, the entropy defect ($S_\text{D}$) quantifies the missing entropy (order) associated with assembling correlated particles. (Taken from Livadiotis & McComas 2024a.)

The entropy defect provides the physical understanding and mathematical expression of how the entropy of a system is shared among its parts. It is formulated as follows. Let two, originally independent, subsystems A and B with entropies $S_\text{A}$ and $S_\text{B}$, respectively, be assembled into a combined system of entropy $S_{\text{A} \oplus \text{B}}$. The order induced by the developed additional correlations causes the system's entropy to decrease, $S_{\text{A} \oplus \text{B}} - (S_\text{A} + S_\text{B}) < 0$, with the missing entropy defining the entropy defect $S_\text{D} \equiv (S_\text{A} + S_\text{B}) - S_{\text{A} \oplus \text{B}}$. Then, the partitioning of entropies is formulated by the addition-rule of entropies, called kappa-addition, that is,

$$S_{\text{A} \oplus \text{B}} = S_\text{A} + S_\text{B} - \tfrac{1}{\kappa} \cdot S_\text{A} \cdot S_\text{B}, \tag{1}$$

(where all entropies are expressed in units of Boltzmann constant, $k_\text{B}$). This generalizes the simple addition of entropies that characterizes the classical Boltzmann-Gibbs (BG) entropy, $S_{\text{A} \oplus \text{B}} = S_\text{A} + S_\text{B}$; indeed, the classical framework of BG statistical mechanics is consistent with thermodynamics in the absence of particle correlations, i.e., when $\kappa \to \infty$. Further, it has been shown that the entropy following this addition rule is given by Tsallis entropic formulation (e.g., Tsallis 1988; Enciso & Tempesta 2017; Livadiotis 2018a,b; Livadiotis & McComas 2024a, Suppl A), whose maximization within the context of canonical ensemble leads to the kappa distributions (Livadiotis & McComas 2009; 2024b; Livadiotis 2014). Therefore, the triplet of entropy summation, BG entropy, and MB distribution,

$$\begin{aligned} S_{\text{A} \oplus \text{B}} = S_\text{A} + S_\text{B} &\Leftrightarrow S = -\int p(\varepsilon) \ln p(\varepsilon) g(\varepsilon) d\varepsilon \\ &\Leftrightarrow p(\varepsilon) \sim \exp(-\varepsilon / k_\text{B} T), \end{aligned} \tag{2a}$$

has been generalized to the addition rule of entropy defect, the Tsallis entropy, and the kappa distribution:

$$S_{A \oplus B} = S_A + S_B - \tfrac{1}{\kappa} \cdot S_A \cdot S_B \Leftrightarrow S = \kappa \cdot \left[1 - \int p(\varepsilon)^{1+1/\kappa} g(\varepsilon) d\varepsilon \right]$$
$$\Leftrightarrow P(\varepsilon) \sim p^{1+\tfrac{1}{\kappa}}(\varepsilon) \sim \left[1 + \frac{\varepsilon}{(\kappa - \tfrac{3}{2})k_B T}\right]^{-\kappa-1}, \quad (2b)$$

where $g(\varepsilon)$ denotes the density of energy states, while the ordinary $p$ and escort $P$ distributions are detailed aspects of nonextensive statistical mechanics (e.g., see: Livadiotis 2017, Ch. 1). While there are new publications on many theoretical developments and applications related to the entropy defect (Table 1), several authors previously also used the same mathematical addition rule, for which the entropy defect has now established the physical foundation, and in fact, thermodynamic requirement.

**Table 1. Entropy Defect: Theory and Applications**

| (A) Entropy of space plasmas | |
|---|---|
| Generalization of Sackur-Tetrode entropy | Livadiotis & McComas 2021 ; 2023physscr |
| Entropy of open systems with correlations | *this paper* |
| (B) Entropy exchange among interacting systems | |
| Transport equation of kappa | Livadiotis & McComas 2023c |
| Pickup Ion thermodynamics | Livadiotis et al. 2024b |
| (C) Thermodynamics (systems in stationary states) | |
| Origin of kappa distributions | Enciso & Tempesta 2017; Livadiotis 2018a; 2018b; Livadiotis & McComas 2022; 2023a |
| Zero-th law | Livadiotis 2018a; Livadiotis & McComas 2023b; d |
| Kinetic and Thermodynamic definitions | Abe 2001; Livadiotis 2018a; Livadiotis & McComas 2021 |
| Superstatistics | Ourabah 2024 |
| (D) Entropy evolution (nonstationary systems) | |
| Entropy addition rule | Abe 2001; Livadiotis 2018a; Livadiotis & McComas 2023a; 2024a |
| Entropic upper boundary | Livadiotis & McComas 2023a; 2024a |
| Stationarity Deviation Index (SDI) | *this paper* |
| (E) Thermodynamic relativity | |
| Connection between entropy & velocity | Livadiotis & McComas 2024a |
| Asynchronous relativity | Livadiotis & McComas 2024a; 2025 |

The entropy defect does not require particles to be kappa distributed, and thus is the basis of characterizing the thermodynamics of nonstationary, as well as, stationary systems. Nonstationary systems do not have kinetically or thermodynamically defined temperatures, but they can be associated with an "equivalent temperature." In contrast, they are still characterized by the thermodynamic parameter kappa, as this emerges from the magnitude of entropy defect (Livadiotis & McComas 2024a). In such cases, there is no stationary distribution and the entropy evolves towards its maximum limit (kappa). Eventually, such systems can evolve into stationary states that stabilize into kappa distributions, where the magnitude kappa of entropy defect also becomes the well-known $\kappa$ that parameterizes kappa distributions.

The stationarity of thermodynamic states is connected with the thermodynamic polytropic processes that characterize the plasma flow. A streamline is defined by the path of an infinitesimal elementary fluid parcel along the flow (e.g., Kartalev et al. 2006). Thermodynamic polytropic processes characterize and formulate the evolution of the population of plasma along streamlines. Indeed, in order correctly observe a polytropic process, one must continuously observe the same population of plasma as it is heated. In other words, the polytropic equation is only valid along a streamline (Newbury et al. 1997; Nicolaou et al. 2014; Nicolaou et al. 2021). Along a streamline with a certain polytropic process, the polytropic relationship and its involved parameters remain invariant (Newbury et al. 1997; Kartalev et al. 2006), that is, (i) the polytropic pressure $\Pi$ and (ii) the polytropic index $\gamma$ (Livadiotis 2016; Livadiotis et al. 2022). This leads to an important property of streamlines: (i) The polytropic pressure is a function of temperature, density, and thermal pressure. The equation of state connects these three thermodynamic variables, so only two are actually independent. Typically, we work with density and temperature, as the independent thermodynamic variables. Therefore, along a streamline with a certain polytropic process, temperature and density vary so that $\Pi$ remains fixed. (ii) The thermodynamic parameter kappa is connected with the polytropic index (in a one-to-one relationship) through the long-range interaction that characterizes the dynamics of the plasma particles (e.g., Livadiotis 2018c; 2019; Nicolaou & Livadiotis 2019). Therefore, along a streamline with a certain polytropic process, the thermodynamic kappa is fixed, and the respective distribution is stationary (kappa distribution).

The polytropic process that characterizes the flowing plasma is not "purely" fixed, however, as there are no truly infinitesimal elementary volumes, and practically speaking, the evolved plasma population follows a bundle of neighboring streamlines. Therefore, along its flow, plasma can be characterized by a fluctuating value of kappa. While the average kappa is about the same in a plasma that resides in a stationary state, measures of its fluctuations, such as its standard deviation must play an important role in the transition of a system towards nonstationarity.

Deviations (either random or systematic) of the system from a stationary state quantify its natural tendency to leave stationarity. However, we do not currently have an actual measure of this natural tendency to leave stationarity – the purpose of this paper is to develop such a measure. For this, we need once again the entropy defect, because we deal with systems residing in stationary states (thus, kappa distributed) and departing from stationary states (thus, their entropy evolves). Using the entropy defect, we will construct the entropy of open particle systems, such as, space plasmas. The formulation will lead to a surprising entropic inequality that characterizes small deviations of particles along plasma flow streamlines.

Let $S$ be the entropy of the particle system residing in stationary states with specific thermodynamic parameters, kappa $\kappa$, temperature $T$, and density $n$. If there were no correlations, the entropy would have been given by its classical value, $S_\infty$ (with infinity noting $\kappa\rightarrow\infty$). The entropy dependence on $T$ and $n$ is enclosed in the form of $S_\infty$, thus the entropy can be written in the convenient way of $S = S(S_\infty,\kappa)$. The plasma flow is characterized by a certain fixed value of kappa, such that only $S_\infty$, and therefore the entropy $S$ varies.

In this study, we find that the entropy varies so that the variations of $S_\infty$ are less than twice the kappa, or $\Delta S_\infty/\kappa \leq 2$. We show that this entropic deviation ratio scales with the kappa deviation, $\Delta\kappa$, as the system fluctuates from the one stationary state to another. In fact, the exponent $\mu$ in the power-law scaling of $\Delta S_\infty / \kappa \sim \Delta\kappa^\mu$ or $\mu = \lim_{\Delta\kappa \to 0} \ln(\Delta S_\infty / \kappa) / \ln(\Delta\kappa)$ measures the natural tendency of the system to depart from stationarity and we call this measure the stationarity deviation index (*SDI*).

The paper is organized as follows. In Section 2 we use the entropy defect and the respective addition rule of entropies to construct the entropy formulation of closed and open systems. In Section 3, we show the statistics of entropy, and derive the inequality of entropic deviation ratios, $\Delta S_\infty/\kappa \leq 2$. In Section 4, we show that this inequality is observed in heliospheric plasma data from three separate missions. In the discussion in Section 5, we demonstrate the property of the entropic deviation ratio to characterize the stationarity of the system, and its natural tendency to become nonstationary. Finally, the conclusions are provided in section 6.

## 2. Entropy of systems with correlations
### 2.1. Algebra of entropies

The kappa addition between the entropies of A and B, is denoted by the symbol $\oplus$,

$$S_{A \oplus B} \equiv S_A \oplus S_B = S_A + S_B - \tfrac{1}{\kappa} \cdot S_A \cdot S_B, \qquad (3)$$

where its algebra was fully developed by Livadiotis & McComas (2023epl). These authors showed that a negative amount of entropy $S$ is not given by using opposite sign, $-S$, because the addition rule of entropies is not that simply given by the classical BG. In general, the negative of a term is determined by their sum being equal to zero. Namely, given the entropy $S$, its negative value, $\bar{S}$, is given by $0 = S \oplus \bar{S} = S + \bar{S} - \tfrac{1}{\kappa} \cdot S \cdot \bar{S}$, that is, $\bar{S} = -S / (1 - \tfrac{1}{\kappa} S)$.

Likewise, the kappa-subtraction of two entropic terms, B from A, denoted with $S_{A \oplus \bar{B}}$, is determined by the kappa addition of A and the negative B, denoted with $\bar{B}$. Hence, adding $S_{A \oplus \bar{B}}$ on the entropy of B, would end up with the entropy of A, $S_A = S_{A \oplus \bar{B}} \oplus S_B = S_{A \oplus \bar{B}} + S_B - \tfrac{1}{\kappa} \cdot S_{A \oplus \bar{B}} \cdot S_B$, leading to

$$S_{A \oplus \bar{B}} = \frac{S_A - S_B}{1 - \tfrac{1}{\kappa} S_B}. \qquad (4)$$

### 2.2. Closed systems

We wish to derive the entropy of a system characterized by correlations among its particles, whose measure is given by the value of inverse kappa, $1/\kappa$ (Abe 1999; Asgarani & Mirza 2007; Livadiotis & McComas 2011a; 2021; 2023a; Livadiotis 2015b; Nelson et al. 2017; Livadiotis et al. 2021). For this, we first start with the case of systems with a fixed number of particles $N$, and find how the entropy of a system shifts from $S_n$ to $S_{n+1}$ after absorbing an elementary entropy $\sigma$, that is,

$$S_{n+1} = S_n + \sigma - \tfrac{1}{\kappa} \cdot S_n \cdot \sigma, \tag{5}$$

which is more conveniently written in the product form:

$$1 - \tfrac{1}{\kappa} S_{n+1} = (1 - \tfrac{1}{\kappa} S_n) \cdot (1 - \tfrac{1}{\kappa} \sigma). \tag{6}$$

Then, by induction after $N$ iterations,

$$1 - \tfrac{1}{\kappa} S_n = (1 - \tfrac{1}{\kappa} S_{n-1}) \cdot (1 - \tfrac{1}{\kappa} \sigma) = (1 - \tfrac{1}{\kappa} S_{n-2}) \cdot (1 - \tfrac{1}{\kappa} \sigma)^2 = \cdots = (1 - \tfrac{1}{\kappa} S_0) \cdot (1 - \tfrac{1}{\kappa} \sigma)^n. \tag{7}$$

Setting the final entropy, that is, the entropy $S \equiv S_N$ that combines $N$ elementary entropies, we have

$$\frac{1 - \tfrac{1}{\kappa} S}{1 - \tfrac{1}{\kappa} S_0} = \left(1 - \tfrac{1}{\kappa} \sigma\right)^N = \left(1 - \tfrac{1}{\kappa} \frac{S_\infty}{N}\right)^N \xrightarrow{N \gg 1} e^{-\tfrac{1}{\kappa} S_\infty}, \tag{8}$$

where $S_\infty = \lim_{\kappa \to \infty} = N \cdot \sigma$ stands for the entropy of the system if there were no correlations ($\kappa \to \infty$). Then, for no initial entropy, $S_0 = 0$, we end up with

$$S = \kappa \cdot \left(1 - e^{-\tfrac{1}{\kappa} S_\infty}\right). \tag{9}$$

*2.3. Open systems*

We continue with the same approach as in the previous section, but now we allow either an addition or subtraction of the elementary entropy; then, we proceed to the statistical distribution of their difference, and its expectation value. Elementary entropies may be added or subtracted from the open system. For additions, Eq.(5) still holds, while for subtractions, Eq.(4) gives

$$S_{n+1} = \frac{S_n - \sigma}{1 - \tfrac{1}{\kappa} \sigma}, \tag{10}$$

rewritten in the product form

$$1 - \tfrac{1}{\kappa} S_{n+1} = (1 - \tfrac{1}{\kappa} S_n) \cdot \left(1 - \tfrac{1}{\kappa} \sigma\right)^{-1}, \tag{11}$$

or combining both product forms for adding (6) and subtracting (11),

$$1 - \tfrac{1}{\kappa} S_{n+1} = (1 - \tfrac{1}{\kappa} S_n) \cdot \left(1 - \tfrac{1}{\kappa} \sigma\right)^{\pm 1} \tag{12}$$

Now we let $N_+$ and $N_-$ denote the total number of elementary entropies added and subtracted, respectively, to/from the system (Figure 2), i.e.,

$$\frac{1 - \tfrac{1}{\kappa} S}{1 - \tfrac{1}{\kappa} S_0} = \left(1 - \tfrac{1}{\kappa} \sigma\right)^{N_+ - N_-} \tag{13}$$

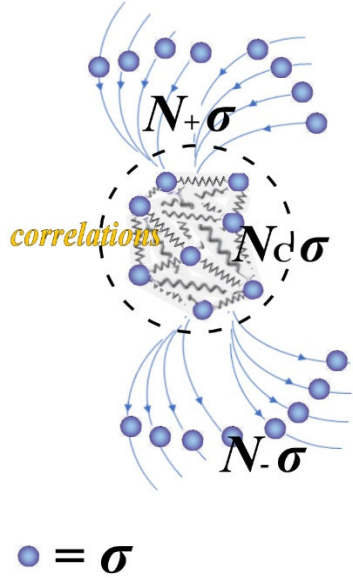

**Figure 2.** Flow of $N_+$ entering and $N_-$ exiting elementary entropies, while the system is collecting the number of correlated elementary entropies, $N_C = N_+ - N_-$.

We estimate the number $N_C \equiv N_+ - N_-$ of the correlated particles that are collected in the open system. This is a random variable, which is noted with $m \equiv N_+ - N_-$ for the following derivations. This may have any positive or negative integer value, but with different probability assigned to each $m$, noted with $P_m$, with normalization $\sum_{m=-\infty}^{+\infty} P_m = 1$. The physical meaning of negative $m$ values is that the exiting number of elementary entropies is larger than the respective entering number; this is physically reasonable once the contribution of all negative $m$ values is less than the contribution of all positive $m$ values, i.e., $\sum_{m=1}^{+\infty} m P_m > \sum_{m=-\infty}^{-1} |m| P_m$ or $\sum_{m=1}^{+\infty} m P_m > \sum_{m=1}^{+\infty} m P_{-m}$, thus $\langle m \rangle = \sum_{m=-\infty}^{+\infty} m P_m > 0$, or $\langle N_+ \rangle > \langle N_- \rangle$. Certainly, if there exists an initial number of elementary entropies, $(N_+ - N_-)_0$, then, finite non-positive values of $\langle m \rangle$, i.e., $\langle N_+ \rangle - \langle N_- \rangle \leq 0$ is also allowed, as long as $(N_+ - N_-)_0 + \langle N_+ \rangle - \langle N_- \rangle \geq 0$.

The respective probability distribution is given by the Skellam distribution, $P_m^{\mathrm{sk}}$, that is, the difference of two independent Poisson distributions (Skellam 1946),

$$P_m^{\mathrm{sk}} = e^{-\langle N_+ \rangle - \langle N_- \rangle} \cdot \left( \frac{\langle N_+ \rangle}{\langle N_- \rangle} \right)^{\frac{1}{2}m} \cdot I_m\left(2\sqrt{\langle N_+ \rangle \langle N_- \rangle}\right), \quad (14)$$

which involves the average entering $\langle N_+ \rangle$ and exiting $\langle N_- \rangle$ elementary entropies, and the modified Bessel function of the first type $I_m$. Then, the entropy is given by

$$\frac{1 - \tfrac{1}{\kappa} S}{1 - \tfrac{1}{\kappa} S_0} = \sum_{m=-\infty}^{\infty} P_m^{\mathrm{sk}} \cdot \left(1 - \tfrac{1}{\kappa}\sigma\right)^m. \quad (15)$$

Note that the normalization of the probability reaches the infinite limiting values of $m$, i.e., $-\infty < m < +\infty$. This may appear unphysical but it simply means that very large numbers can practically be written as infinity. The same sort of normalization of the probability applies to other physical variables. For example, the energy probability distributions are normalized over energies $0 \leq E < +\infty$, even though infinite energy is surely unphysical.

Then, substituting the probability distribution shown in Eq.(14) in Eq.(15), we obtain:

$$\frac{1-\frac{1}{\kappa}S}{1-\frac{1}{\kappa}S_0} = e^{-\langle N_+\rangle - \langle N_-\rangle} \cdot \sum_{m=-\infty}^{\infty}\left[\left(1-\frac{1}{\kappa}\sigma\right)\left(\frac{\langle N_+\rangle}{\langle N_-\rangle}\right)^{\frac{1}{2}}\right]^m \cdot I_m\left(2\sqrt{\langle N_+\rangle\langle N_-\rangle}\right). \tag{16}$$

Also, from the probability generating function of the Skellam distribution (e.g., Abramowitz & Stegun 1972; Karlis & Ntzoufras 2008; Tathe & Ghosh 2024), we get

$$\sum_{m=-\infty}^{\infty}\left[\left(1-\frac{1}{\kappa}\sigma\right)\left(\frac{\langle N_+\rangle}{\langle N_-\rangle}\right)^{\frac{1}{2}}\right]^m \cdot I_m\left(2\sqrt{\langle N_+\rangle\langle N_-\rangle}\right) = e^{\langle N_+\rangle\left(1-\frac{1}{\kappa}\sigma\right)^{+1} + \langle N_-\rangle\left(1-\frac{1}{\kappa}\sigma\right)^{-1}}, \tag{17}$$

thus,

$$\frac{1-\frac{1}{\kappa}S}{1-\frac{1}{\kappa}S_0} = e^{-\langle N_+\rangle\left(\frac{1}{\kappa}\sigma\right) + \langle N_-\rangle\left[\left(1-\frac{1}{\kappa}\sigma\right)^{-1}-1\right]} = e^{-(\langle N_+\rangle - \langle N_-\rangle)\frac{1}{\kappa}\sigma} \cdot e^{\langle N_-\rangle\cdot\frac{(\frac{1}{\kappa}\sigma)^2}{1-\frac{1}{\kappa}\sigma}} = e^{-(\langle N_+\rangle - \langle N_-\rangle)\cdot\frac{1}{\kappa}\sigma\left(1-\frac{\langle N_-\rangle}{\langle N_+\rangle - \langle N_-\rangle}\cdot\frac{\frac{1}{\kappa}\sigma}{1-\frac{1}{\kappa}\sigma}\right)}. \tag{18}$$

Setting the initial entropy to zero, we have

$$\tfrac{1}{\kappa}S = F(\tfrac{1}{\kappa}\sigma), \text{ with } F(x) \equiv 1 - e^{-N_C\cdot x\left(1-\nu\cdot\frac{x}{1-x}\right)}, \tag{19}$$

where we denote the exiting fraction $\nu = \langle N_-\rangle / \langle N_C\rangle$. The number of the collecting elementary entropies that form the entire correlated system is given by the average $\langle N_C\rangle \equiv \langle N_+\rangle - \langle N_-\rangle$, thus, $\langle N_-\rangle = \langle N_+\rangle \cdot \nu / (\nu+1)$. Figure 3 plots function $F$ shown in Eq.(19), that is, the entropy $S$ as a function of elementary entropy, both normalized by kappa.

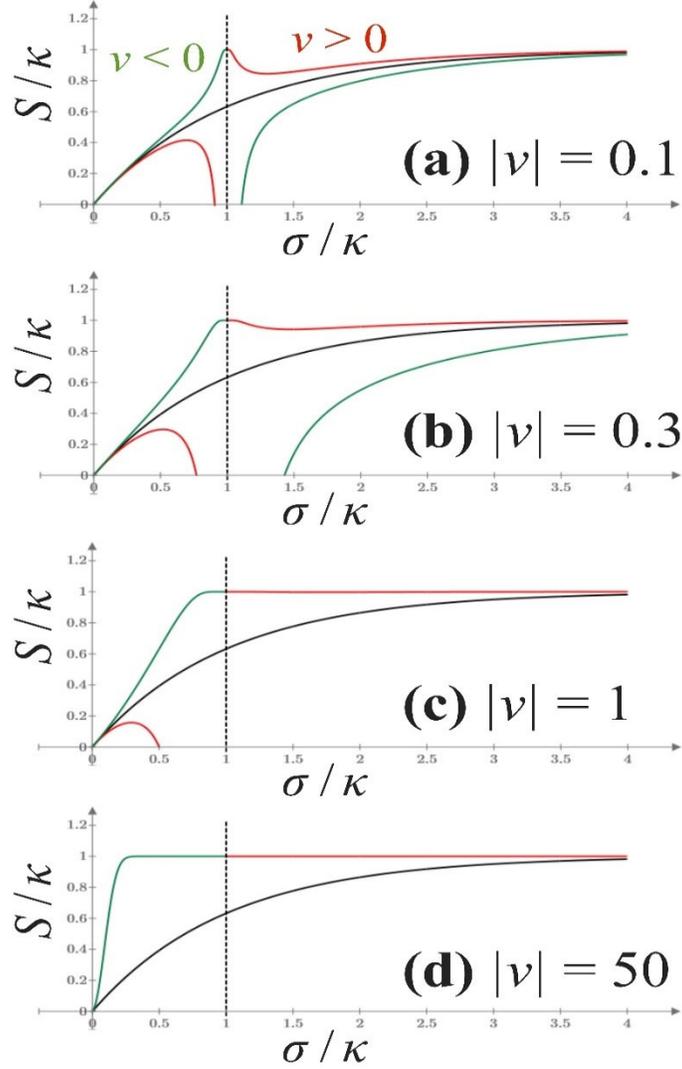

**Figure 3.** Plot of entropy $S$ as a function of the elementary entropy $\sigma$. Both entropies are expressed as ratios with kappa, as shown in Eq.(19). The parameter $v$ indicates the fraction of the average exiting $\langle N_- \rangle$ per collecting $\langle N_C \rangle \equiv \langle N_+ \rangle - \langle N_- \rangle$ elementary entropies. We consider a general case where the average numbers can be either $\langle N_+ \rangle > \langle N_- \rangle$ or $\langle N_+ \rangle \leq \langle N_- \rangle$, thus the parameter $v$ can be either positive (red) or negative (green). Each panel shows plots with different values of $|v|$; for comparison, we also plot the case of zero exiting elementary entropies, $v=0$, (black), which corresponds to the closed system's entropy, Eq.(9).

Next, we write the entropy in terms of the BG entropy,

$$S_\infty = \left( \langle N_+ \rangle - \langle N_- \rangle \right) \cdot \sigma, \tag{20}$$

thus,

$$S = \kappa \cdot \left[ 1 - e^{-\frac{1}{\kappa} S_\infty \cdot \left( 1 - v \cdot \frac{\frac{1}{\kappa} S_\infty}{N_C - \frac{1}{\kappa} S_\infty} \right)} \right], \tag{21}$$

or

$$S = \kappa \cdot \left[ 1 - e^{-\frac{1}{\kappa}(S_\infty - \delta S)} \right], \tag{22a}$$

where the corresponding BG entropy is characterized by a negative entropic deviation,

$$\delta S = \langle N_- \rangle \cdot \sigma^2 / (\kappa - \sigma) \cong \langle N_- \rangle \cdot \tfrac{1}{\kappa} \sigma^2 . \tag{22b}$$

This entropy deviation is released in addition to the $\langle N_- \rangle$ elementary entropies, i.e., $\langle N_- \rangle \cdot \sigma$, and corresponds to the particle decorrelations. Indeed, the decorrelation released entropy per each exiting elementary entropy, $\delta S / \langle N_- \rangle$, is $\sim \tfrac{1}{\kappa}\sigma^2$; this is shown by the addition of two elementary entropies, $\sigma \oplus \sigma = \sigma + \sigma - \tfrac{1}{\kappa}\sigma^2$, where the entropy defect caused by the additional correlations is exactly equal to the amount of $\tfrac{1}{\kappa}\sigma^2$.

## 3. Statistics of entropy

The previous developments lead to some interesting findings related to the BG entropy, $S_\infty$. This can be especially useful as the BG entropy is a simple function of the plasma moments (temperature and density), which are routinely measured in space plasmas. The amount of the entropy that comes from the decorrelation is subtracted from the elementary entropies that, on average, stay in the combined system, $S_\infty \geq \delta S$; this is equivalent to having nonnegative combined entropy $S \geq 0$ (as shown in Eq.(19)). This inequality leads to:

$$\tfrac{1}{\kappa}\sigma \leq \frac{\langle N_+ \rangle - \langle N_- \rangle}{\langle N_+ \rangle}, \tag{23a}$$

namely, the ratio of the elementary entropy per entropy's maximum value, $\kappa$, is limited by the difference between the average entering $\langle N_+ \rangle$ and exiting $\langle N_- \rangle$ number of elementary entropies, normalized to the entering one. Then, the respective BG entropy is limited by

$$\tfrac{1}{\kappa} S_\infty \leq \frac{(\langle N_+ \rangle - \langle N_- \rangle)^2}{\langle N_+ \rangle}. \tag{23b}$$

Even more interesting is the deviation of BG entropy,

$$\tfrac{1}{\kappa}\Delta S_\infty = (\langle N_+ \rangle - \langle N_- \rangle) \cdot \tfrac{1}{\kappa}\sigma \leq \frac{\Delta(\langle N_+ \rangle - \langle N_- \rangle)^2}{\langle N_+ \rangle} \sim \frac{\sigma_{\text{SK}}^2}{\langle N_+ \rangle} = \frac{\langle N_+ \rangle + \langle N_- \rangle}{\langle N_+ \rangle}, \tag{24}$$

where the deviation of the square difference $\Delta(\langle N_+ \rangle - \langle N_- \rangle)^2$ is approached by the variance of $N_+ - N_-$, that is, the variance of Skellam distribution, $\sigma_{\text{SK}}^2$, which is equal to $\langle N_+ \rangle + \langle N_- \rangle$.

On the other hand, the average exiting number of elementary entropies cannot be larger than respective average entering ones, i.e., $\langle N_- \rangle \leq \langle N_+ \rangle$ or $\langle N_- \rangle / \langle N_+ \rangle \leq 1$. Hence, the deviation of BG entropy follows the inequality:

$$\tfrac{1}{\kappa}\Delta S_\infty \leq 2. \tag{25}$$

The thermodynamic kappa depends on the number $N_C$ of the correlated particles (e.g., Livadiotis & McComas 2014; Livadiotis et al. 2018). In particular, it equals the half dof ($D$) carried by each correlated particle, so that the total correlated dof are $D \cdot N_C$. The dof per correlated element (here, the elementary entropy $\sigma$), noted with $D$, are not given by just the sum of all the kinetic dof, $d_K$. There is also a part of it that depends only on the correlations (e.g., their type, strength, etc.), called internal dof and denoted with $d_0$; thus, $D = d_0 + d_K$. Then, when we refer to one cluster of correlated constituents (e.g., here, elementary entropies), the kappa equals $\kappa = \frac{1}{2}d_0 + \frac{1}{2}d_K N_C$. Of course, if the system's size is extended to $N$ clusters of $N_C$ elementary entropies each, the kappa equals $\kappa = \frac{1}{2}d_0 + \frac{1}{2}d_K N_C \cdot N$. In a cluster of correlated constituents, we express the kappa per correlated constituent, $\kappa(1) \equiv \kappa / N_C = \frac{1}{2}d_0 / N_C + \frac{1}{2}d_K$, but the internal dof depends on correlations, $d_0 = d_0(N_C)$. Then, the invariant kappa is defined by: $\kappa_0 \equiv \frac{1}{2}d_0(N_C)/N_C$, so that $\kappa(1) = \kappa_0 + \frac{1}{2}d_K$. Since the BG entropy is extensive, $\Delta S_\infty = N_C \cdot \Delta S_\infty(1)$, or

$$\frac{1}{\kappa}\Delta S_\infty = \frac{1}{\kappa(1)}\Delta S_\infty(1) \leq 2. \qquad (26)$$

**4. Applications for heliospheric plasma protons**

*4.1. Data*

We use three datasets, observed in three different space plasmas from vastly different locations in the heliosphere:

(1) Solar energetic particle (SEP) event, recorded by the Integrated Science Investigation of the Sun instrument (ISOIS; McComas et al. 2016) onboard Parker Solar Probe (PSP; Fox et al. 2016), near 0.35 au on 2022-02-15; in general, SEP observations provide their flux time series over a broad range of energies (e.g., Kouloumvakos et al. 2019; McComas et al. 2019; Desai et al. 2020; Joyce et al. 2020; Mitchell et al. 2020; Schwadron et al. 2020; Wiedenbeck et al. 2020; Cohen et al. 2021, 2024; Giacalone et al. 2021; Dresing et al. 2023; Khoo et al. 2024; Palmerio et al. 2024); the examined event was analyzed by Cuesta et al. (2024; 2025; see also: Cohen et al. 2024; Livadiotis et al. 2024; Sarlis et al. 2024) to derive the thermodynamic variables of the energetic particle protons (density, temperature, and kappa);.

(2) Solar wind protons recorded by the Solar Wind Experiment (SWE) instrument onboard WIND mission near Earth (~1 au) (Ogilvie et al. 1995; see also Wilson et al. 2019); we use the first 72 days of the year 1995 as shown in (Livadiotis & Desai 2016; Livadiotis 2018c; Livadiotis et al. 2020), which includes the density and temperature, while the kappa is derived according to the method of Livadiotis et al. (2018; see also Nicolaou & Livadiotis 2016; Livadiotis 2018c; Nicolaou & Livadiotis 2019).

(3) Plasma protons in the heliosheath, which are the source of energetic neutral atoms (ENAs) recorded by IBEX-Hi onboard Interstellar Boundary Explorer (IBEX) mission (McComas et al. 2009a;b); for the years 2009-2016, data were fitted to kappa distributions by Livadiotis et al. (2022; following the method of Livadiotis et al. 2011) and derived consistent thermodynamic variables of the heliosheath plasma over many directions across the sky (see also: Dayeh et al. 2012; Fuselier et al. 2014).

*4.2. Method*

We use the density and temperature data with the Sackur (1911) - Tetrode (1912) equation for BG entropy (per particle), i.e.,

$$S_\infty = \ln(T^{\frac{1}{2}D} / n) + const., \qquad (27)$$

where we set the dimensionality to *D*=3. We ignore the constant, as we derive the sequential entropic differences elementary entropy, $\Delta S_\infty$. Then, we use the thermodynamic kappa datasets to find the ratios $\Delta S_\infty / \kappa$.

The entropic differences are calculated using datasets with the highest publicly available time resolution. The data used are not actually single point measurements but averages over each time-resolution interval. Then, what we measure are the differences between sequential BG entropies of averaged intervals, whose expectation value is given by the standard deviation.

*4.3. Results*

Figure 3 plots histograms of the values of $\Delta S_\infty / \kappa$, derived from the three datasets. All histograms are well-fitted by a statistical distribution $\sim [1 + (x/\sigma_X)^2]^{-2}$ (Livadiotis 2007), with $x \equiv \Delta S_\infty / \kappa$ and $\sigma_X = 0.1$, 0.17, and 0.45, respectively for the three datasets. Upper limit of all values is $\frac{1}{\kappa}\Delta S_\infty \leq 2$. The peak of all the distributions is at zero entropic deviation ratio, $\Delta S_\infty / \kappa \sim 0$, corresponding to times of fixed entropy, with the system residing in a stationary state. The larger absolute values of the estimated ratio indicate that the system is characterized by some natural tendency to depart from its current stationary state. The strong natural tendency of the probability to be very near zero indicates strong stationarity in all three space plasma populations.

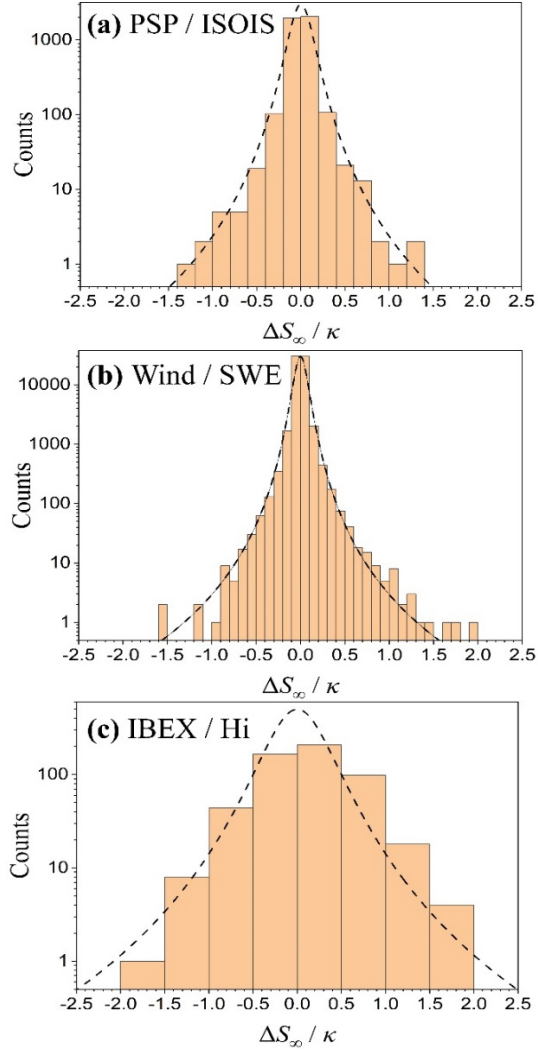

**Figure 4.** Distribution of the values $\Delta S_\infty / \kappa$ in sampling counts, derived from the three datasets (1) SEP event on 2022-02-15, recorded by PSP/ISOIS; (2) solar wind protons recorded by WIND/SWE during the first 72 days of 1995; and (3) plasma protons in the heliosheath, source of energetic neutral atoms (ENAs) recorded by IBEX-Hi between 2009-2016.

## 5. Discussion

In this study, we showed the power-law relationship of the entropic deviation ratio and the kappa deviation, as well as providing a measure of the natural tendency of the system to deviate from stationarity.

### 5.1. Sequential increments of entropic deviation ratio

A polytropic process is the variation of thermodynamic parameters as the plasma particles flow along streamlines according to a polytropic relationship. Kappa distributions and their associated entropy are strongly connected with polytropic processes. Variations of entropy occurs while the flow fluctuates along neighboring streamlines, thus the entropic deviation ratio is connected with variation of polytropic parameters. In order to understand this, we first review the connection of polytropic behavior with kappa distributions.

The thermodynamic phase diagram plots two independent parameters, typically, thermal pressure vs. temperature ($T,P$) or density vs. temperature ($T,n$). The polytropic relationship converts the one parameter (pressure or density) with the polytropic pressure (Livadiotis 2016; Livadiotis et al. 2022). This is the constant included in the polytropic behavior, i.e., $P \cdot n^{-\gamma} = const. \equiv \Pi$. It is often neglected, as we typically care only about the polytropic power-law relationship, e.g., $P \propto n^{\gamma}$, instead of the magnitude of this constant. In an isobaric process, the polytropic index is zero and the pressure constant. For any nonzero polytropic index, the pressure does not remain constant, but the polytropic pressure does. The polytropic pressure generalizes the role of the thermal pressure in an isobaric process, providing that thermodynamic variable that remains invariant in any non-isobaric process.

Along a streamline both polytropic index $\gamma$ and pressure $\Pi$ remain constant. Then, the polytropic process is described by

$$P = \Pi \cdot n^{\gamma} \text{ or } n = \Pi^{-\nu} \cdot T^{\nu}, \qquad (28)$$

where the polytropic index $\gamma$ is connected with the exponent $\nu$, often used as the secondary polytropic index,

$$\gamma = 1 + 1/\nu \iff \nu = 1/(\gamma - 1). \qquad (29)$$

The description of thermodynamic parameters with polytropic processes is one-to-one with the description of particle energies with kappa distributions (Livadiotis 2019). This connection is caused by local long-range interactions, responsible for the existence of particle correlations that lead to the thermodynamics of kappa distributions (Livadiotis & McComas 2011a; 2022), as well as actuating local polytropic processes (Livadiotis 2015c, 2018c; 2019). As a result, the thermodynamic kappa is connected with the polytropic index, such as, $\nu \sim \tfrac{1}{2} - \kappa + D/b$, where D is the dimensionality and $b$ is the radial exponent in potential energy, $\sim 1/r^b$. For weak interactions the last term can be ignored; hence, we write the polytropic expression as

$$\ln \Pi \sim \ln n + (\kappa - \tfrac{1}{2}) \cdot \ln T. \qquad (30)$$

Variations of the polytropic pressure $\Pi$ means the system changes streamlines, e.g., it is fluctuating along nearby streamlines, but still, it is characterized by the same stationary state. However, variations of the polytropic index $\gamma$ (or $\nu$) means the system changes stationary states (e.g., Dayeh & Livadiotis 2022; Katsavrias et al. 2024; Livadiotis & McComas 2010b; 2012). In reality, when a system is changing its stationary state, then both variations of polytropic index and of polytropic pressure are likely occurring.

Therefore, the total polytropic variance $\sigma_{tot}^2$ should include both the variation of polytropic pressure, $\sigma_{\ln \Pi}^2 = \sigma_{\ln n}^2 + (\kappa - \tfrac{1}{2})^2 \cdot \sigma_{\ln T}^2$ and the variance of kappa, $\sigma_{\kappa}^2 \cdot \ln^2 T$, i.e.,

$$\sigma_{\text{tot}}^2 = \sigma_{\ln \Pi}^2 + \sigma_\kappa^2 \cdot \ln^2 T = \sigma_{\ln n}^2 + (\kappa - \tfrac{1}{2})^2 \cdot \sigma_{\ln T}^2 + \sigma_\kappa^2 \cdot \ln^2 T \ . \tag{31}$$

This variation is a result of the system's departing from a stationary state and possibly becoming nonstationary. Since the entropic deviation ratio $\Delta S_\infty / \kappa$ provides a measure of the natural tendency of the system to become nonstationary, there must be positive correlation between the entropic ratio and the total polytropic variation. Such a behavior is shown in Figure 5, where we plot the evolution of several cases of sequential increasing values of the entropic deviation ratio, $\Delta S_\infty / \kappa$ (for the SEP event on 2022-02-15 shown in Figure 4), which exhibit linear increasing trend with the total polytropic variance, $\sigma_{\text{tot}}^2$.

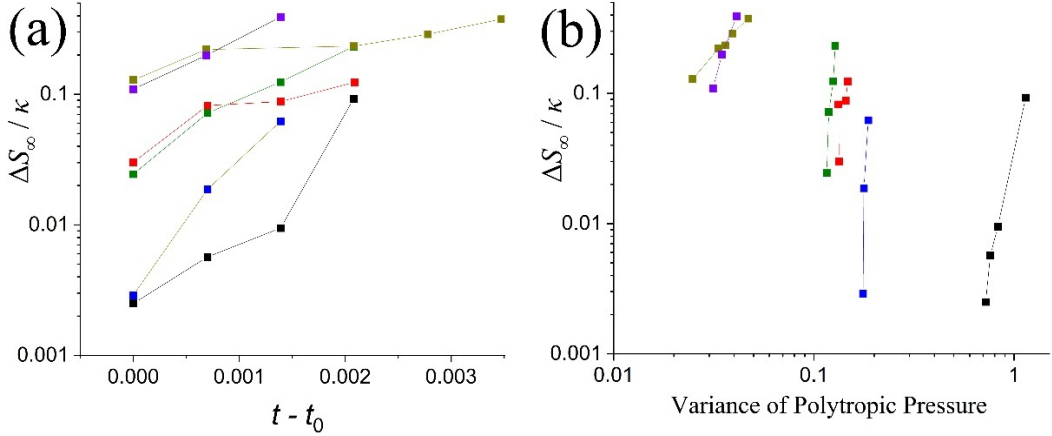

**Figure 5.** (a) Sequential increasing values of $\Delta S_\infty / \kappa$ (for the SEP event on 2022-02-15 shown in Figure 4), plotted with respect to time $t$–$t_0$, with $t_0$ indicating the starting time of the sequential increasing entropies. (b) The values of $\Delta S_\infty / \kappa$ are now plotted with respect to the variance of polytropic pressure, $\sigma_{\text{tot}}^2$, showing a positive correlation between the two parameters.

*5.2. Relationship between entropic deviation ratio and kappa deviation*

Here we calculate, and examine together, the sequential kappa deviations $\Delta \kappa$ and entropic deviation ratios $\Delta S_\infty / \kappa$, determined from the three datasets of (i) PSP/ISOIS, (ii) WIND/SWE, and (iii) IBEX/Hi, described in Section 4.1.

Figure 6 gives our analysis of the SEP event observed by PSP/ISOIS. Figure 6(a) plots the histogram of the sequential kappa deviations, $\log(\Delta \kappa)$. Given this sampling distribution of the different kappa deviations, it is not surprising that the sequential entropic deviation ratios $\log(\Delta S_\infty / \kappa)$ versus the corresponding kappa deviations $\log(\Delta \kappa)$, plotted in Figure 6(b), also shows similar maximum, corresponding to the peak observed in Figure 6(a). Therefore, we normalize the 2D histogram to investigate the actual relationship between the plotted parameters, and thus, we divide the values from Figure 6(b) by those of Figure 6(a) to produce Figure 6(c). This plot shows the 2D histogram of $\log(\Delta S_\infty / \kappa)$ vs. $\log(\Delta \kappa)$, normalized by the 1D histogram of $\log(\Delta \kappa)$, and clearly demonstrates the positive correlation between the two values. (With regard to the construction of normalized 2D histograms, see also Livadiotis & Desai 2016; Livadiotis et al. 2023.) Figure 6(d) plots the weighted mean and standard error of the values of $\log(\Delta S_\infty / \kappa)$, estimated for each of the $\log(\Delta \kappa)$-bins, on which we fit a linear model.

Figure 7 and 8 repeats the previous steps in Figure 6 using the Wind/SWE and IBEX/Hi, respectively. The linear fitting of respective panels (d) provide the value of *SDI*.

The linear regression determines the power-law scaling, $\Delta S_\infty / \kappa \sim \Delta \kappa^{SDI}$, or $\log(\Delta S_\infty / \kappa) \cong$ Intercept $+ SDI \cdot \log(\Delta \kappa)$. The index *SDI* measures the natural tendency of the system to effectively depart from stationarity,

$$SDI \equiv \lim_{\Delta \kappa \to 0} \ln(\Delta S_\infty / \kappa) / \ln(\Delta \kappa). \quad (32)$$

The intercept is a good indicator of the average value of $\log(\Delta S_\infty / \kappa)$. This indicates the value of the entropic deviation ratios $\log(\Delta S_\infty / \kappa)$ if there were no dependence on $\Delta \kappa$ (*SDI* ~ 0), that is, no tendency for the measure of stationarity to increase.

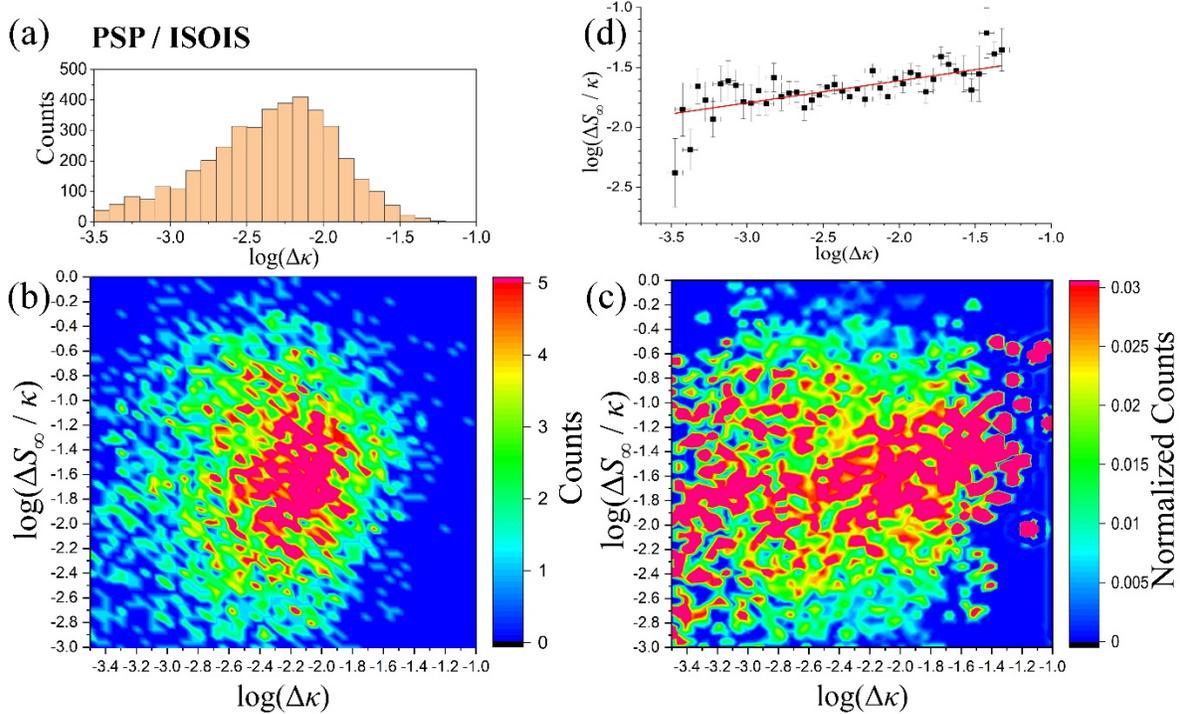

**Figure 6.** (a) Histogram of the sequential kappa deviations, $\log(\Delta \kappa)$, corresponding to the SEP event observed by PSP/ISOIS and described in Section 4.1. (b) 2D histogram of the sequential entropic deviation ratios $\log(\Delta S_\infty / \kappa)$ versus the respective kappa deviations $\log(\Delta \kappa)$. (c) 2D distribution from panel (b), but normalized by the 1-D plotted in (a). Panel (d) shows (i) the weighted mean and standard error of $\log(\Delta S_\infty / \kappa)$ for each of the binned values of $\log(\Delta \kappa)$, where we fit a linear model.

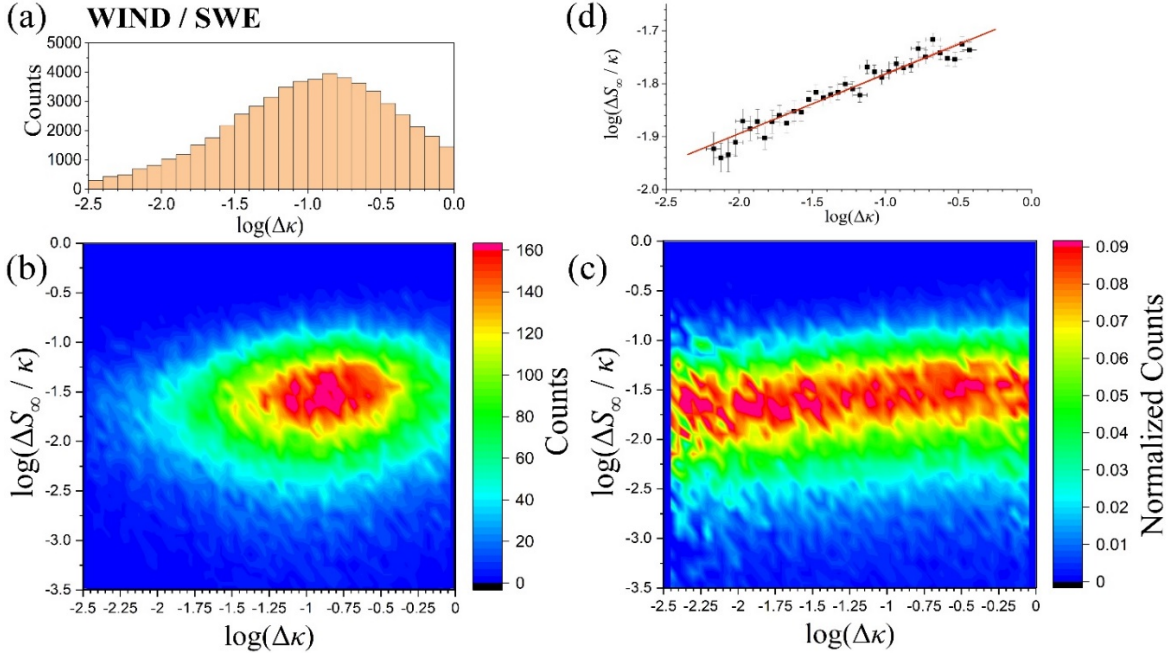

**Figure 7.** Similar to Figure 6, but for the WIND/SWE dataset. (a) 1-D histogram of $\log(\Delta\kappa)$; (b) 2-D histogram of $\log(\Delta S_\infty/\kappa)$ versus $\log(\Delta\kappa)$; (c) normalized 2-D histogram; and (d) weighted mean and standard errors of $\log(\Delta S_\infty/\kappa)$ for each $\log(\Delta\kappa)$-bin.

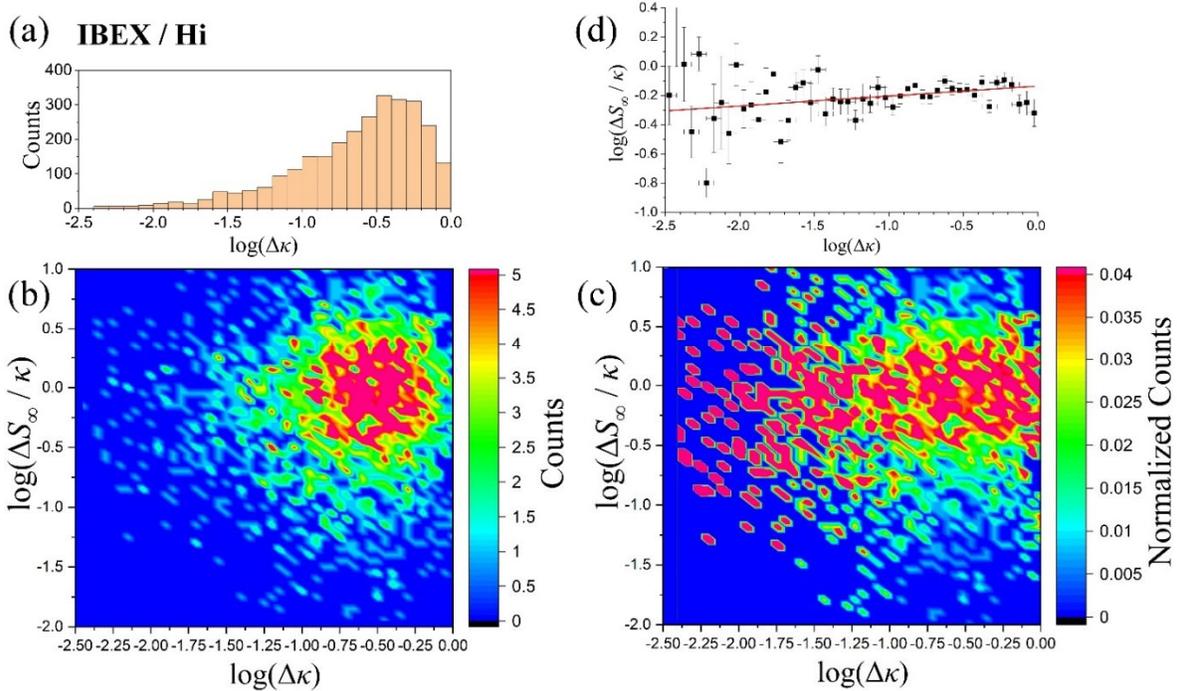

**Figure 8.** Similar to Figure 6, but for the IBEX/Hi dataset. (a) 1-D histogram of $\log(\Delta\kappa)$; (b) 2-D histogram of $\log(\Delta S_\infty/\kappa)$ versus $\log(\Delta\kappa)$; (c) normalized 2-D histogram; and (d) weighted mean and standard errors of $\log(\Delta S_\infty/\kappa)$ for each $\log(\Delta\kappa)$-bin.

**Table 2.** Linear regression and stationarity deviation indices

| Mission | Heliospheric Distance | Intercept | SDI | $\chi^2_{red}$ |
|---|---|---|---|---|
| PSP / ISOIS | 0.35 AU | −1.24 ± 0.07 | 0.19 ± 0.03 | 1.05 |
| WIND / SWE | Near Earth, 1 AU | −1.667 ± 0.007 | 0.114 ± 0.006 | 0.86 |
| IBEX / Hi | Heliosheath, ~ 100 AU | −0.171 ± 0.015 | 0.068 ± 0.028 | 2.05 |

Note: Uncertainties give the standard errors.

Details of the linear regression, and the derived values of Intercept and *SDI* are shown in Table 2. We observe that the stationarity measurements of the examined plasmas have the following comparison:

a) Stationarity: Protons near 1AU > SEPs > Protons in the heliosheath
b) Tendency to leave stationarity: Protons near 1AU < SEPs < Protons in the heliosheath

The examined heliospheric plasmas are characterized by some stability mechanism, avoiding nonstationarity. Indeed, the less stationary the plasma, i.e., the larger values of $\log(\Delta S_\infty / \kappa)$, the smaller their tendency is to become even less stationary, i.e., smaller value of *SDI*.

## 6. Conclusions

This paper used the concept of entropy defect and the respective addition rule of entropies to construct the entropy formulation of open systems, which is critical for developing a measure for stationarity, and its natural tendency to leave stationarity.

The statistical properties of entropy led to the inequality that characterizes the entropic deviation ratios, $\Delta S_\infty/\kappa < 2$. We verified this inequality for three different heliospheric plasma datasets observed from three missions: (1) An SEP event, recorded by PSP/ISOIS on 2022-02-15; (2) Solar wind protons near Earth (~1 au), recorded by WIND/SWE in the first 72 days of the year 1995; and (3) Plasma protons in the heliosheath, source of energetic neutral atoms (ENAs) recorded by IBEX/Hi. Furthermore, we demonstrated the capability of the entropic deviation ratio to characterize the stationarity of the system. We showed that the entropic deviation ratio $\Delta S_\infty / \kappa$ scales as a power-law with the kappa deviation $\Delta\kappa$, as the system fluctuates from the one stationary state to another. In fact, the exponent of this power-law $\lim_{\Delta\kappa \to 0} \ln(\Delta S_\infty / \kappa) / \ln(\Delta\kappa)$ measures the natural tendency of the system to effectively depart from stationarity, so we call it the stationarity deviation index (*SDI*).

The analysis of the above datasets supports that the two developments, the inequality of $\Delta S_\infty / \kappa$, as well as, its power-law relationship with $\Delta\kappa$,

$$\Delta S_\infty / \kappa \leq 2 \text{ and } SDI \equiv \lim_{\Delta\kappa \to 0} \ln(\Delta S_\infty / \kappa) / \ln(\Delta\kappa), \qquad (33)$$

can be used to characterize the stationarity of a system and the magnitude its natural tendency to become nonstationary. Indeed, the value of $\Delta S_\infty / \kappa$ provides a measure of stationarity, while the exponent *SDI* provides a measure of the tendency to leave stationarity. Using the three above examined heliospheric plasmas, we found that the less stationary a plasma is, i.e., the larger values of $\log(\Delta S_\infty / \kappa)$, the smaller its tendency is to become even less stationary, i.e., smaller value of *SDI*.

With the results provided in this study, it is straightforward to consistently characterizing stationarity through the values of entropic deviation ratio and *SDI*, determined from various space physics observations, models, and theoretical studies that use kappa distributions. Future work may investigate the involvement and role of these terms in space plasmas processes. For instance, whether they: (i) are related with the time series complexity of space plasma moments, (ii) can detect nonstationarity, (iii) can forecast SEP events due to sudden increase of their values, and (iv) "feel" shock-related or other discontinuities in the plasma flow and the involved polytropic processes.

Thus, the full strength and capability of the entropic deviation ratio and *SDI* can now be used by the space physics community for analyzing and characterizing the stationarity of space plasmas and by other researchers, who can similarly use them for all other systems characterized by kappa distributions and the stationary states that they represent.

**Acknowledgements**
This work was funded in part by the IBEX mission as part of NASA's Explorer Program (80NSSC18K0237), the IMAP mission as a part of NASA's Solar Terrestrial Probes (STP) Program (80GSFC19C0027), the PSP GI grant 80NSSC21K1767, and the PSP/ISOIS grant NNN06AA01C.

**References**

Abe, S. 1999, Physica A, 269, 403
Abe, S. 2001, Phys. Rev. E, 63, 061105
Abramowitz, M., & Stegun, I. 1972, Handbook of mathematical functions with formulas, graphs, and mathematical tables, National Bureau of Standards Applied Mathematics Series.
Asgarani, S., & Mirza, B. 2007, PhysA, 379, 513
Beck, C., & Cohen, E. G. D. 2003, PhysA, 322, 267
Benetti, M. H., Silveira, F. E. M., & Caldas, I. L. 2023, PhRvE, 107, 055212
Bian, N., Emslie, G. A., Stackhouse, D. J., & Kontar, E. P. 2014, ApJ, 796, 142
Binsack, J. H. 1966, Plasma Studies with the IMP-2 Satellite. PhD Thesis MIT, Cambridge, USA
Boltzmann, L. 1866, Wiener Berichte, 53, 195
Cohen, C. M. S., Christian, E. R., Cummings, A. C., et al. 2021, A&A, 650, A23
Cohen, C. M. S., Leske, R. A., Christian, E. R., et al. 2024, ApJ, 966, 148
Cuesta, M.E., Cummings, A.T., Livadiotis, G., McComas, D.J., Cohen, C. M.S., Khoo, L.Y. et al. 2024, ApJ, 973, 76
Cuesta, M. E., Khoo, L. Y., Livadiotis, G., Shen, M. M., Szalay, J. R., McComas, D. J., et al. 2025, ApJSS, in press, arXiv:2501.14923
Davis, S., Avaria, G., Bora, B., et al. 2023, PhRvE, 108, 065207
Dayeh, M. A., & Livadiotis, G. 2022, ApJL, 941, L26
Dayeh, M.A., McComas, D. J., Allegrini, F., Majistre, B. De, Desai, M. I., Funsten, H. O., et al. 2012, ApJ, 749, 50
Desai, M. I., Mitchell, D. G., Szalay, J. R., et al. 2020, ApJS, 246, 56
Dresing, N., Rodríguez-García, L., Jebaraj, I. C., et al. 2023, A&A, 674, A105
Enciso, A., & Tempesta, P. 2017, J. Stat. Mech., 2017, 123101
Fisk, L. A., & Gloeckler, G. 2014, JGRA, 119, 8733

Fox, N. J., Velli, M. C., Bale, S. D. et al. 2016, *SSRv*, 204, 7
Fuselier, S. A., Allegrini, F., Bzowski, M., Dayeh, M. A., Desai, M., Funsten, H. O. et al. 2014, ApJ, 784, 89
Giacalone, J., Burgess, D., Bale, S.D., et al. 2021, ApJ, 921, 102
Gibbs, J.W. 1902, Elementary Principles in Statistical Mechanics (New York: Scribner's Sons)
Gravanis, E., Akylas, E., & Livadiotis, G. 2020, EPL, 130, 30005
Joyce, C. J., McComas, D.J., Christian, E. R., et al. 2020, ApJS, 246, 41
Karlis, D., & Ntzoufras, I. 2008, IMA J. Management Math., 20, 133, 2008
Kartalev, M., Dryer, M., Grigorov, K., Stoimenova, E.: 2006, J. Geophys. Res. 111, A10107
Katsavrias, C., Nicolaou, G., & Livadiotis, G. 2024, Astron. & Astrophys., 691, L11
Khoo, L. Y., Sánchez-Cano, B., Lee, C. O., et al. 2024, ApJ, 963, 107
Kouloumvakos, A., Rouillard, A. P., Wu, Y., et al. 2019, ApJ, 876, 80
Leubner, M. P. 2002, Ap&SS, 282, 573
Livadiotis, G. 2007, PhysA, 375, 518
Livadiotis, G. 2014, Entrp, 16, 4290
Livadiotis, G. 2015a, JGRA, 120, 880
Livadiotis, G. 2015b, Entrp, 17, 2062
Livadiotis, G. 2015c, JGRA, 120, 1607
Livadiotis, G. 2016, ApJS, 223, 13
Livadiotis, G. 2017, Kappa Distribution: Theory Applications in Plasmas (1st ed.; Netherlands: Elsevier)
Livadiotis, G. 2018a, EPL, 122, 50001
Livadiotis, G. 2018b, NPGeo, 25, 77
Livadiotis, G. 2018c, JGRA, 123, 1050
Livadiotis, G. 2018d, Entrp, 20, 799
Livadiotis, G. 2019, ApJ, 874, 10
Livadiotis, G., & Desai, M. I. 2016, ApJ, 829, 88
Livadiotis, G., & McComas, D. J. 2009, JGRA, 114, 11105
Livadiotis, G., & McComas, D. J. 2010a, ApJ, 714, 971
Livadiotis, G., & McComas, D. J. 2010b, Phys. Scr., 82, 035003
Livadiotis, G., & McComas, D. J. 2011a, ApJ, 741, 88
Livadiotis, G., & McComas, D. J. 2011b, ApJ, 738, 64
Livadiotis, G., & McComas, D. J. 2012, ApJ, 749, 11
Livadiotis, G., & McComas, D. J. 2013a, SSRv, 75, 183
Livadiotis, G., & McComas, D. J. 2013b, AIP Conf. Proc., 1539, 344
Livadiotis, G., & McComas, D. J. 2014, JPlPh, 80, 341
Livadiotis, G., & McComas, D. J. 2021, Entrp, 23, 1683
Livadiotis, G., & McComas, D. J. 2022, ApJ, 940, 83
Livadiotis, G., & McComas, D. J. 2023a, NatSR, 13, 9033
Livadiotis, G., & McComas, D. J. 2023b, EPL, 144, 21001
Livadiotis, G., & McComas, D. J. 2023c, ApJ, 954, 72
Livadiotis, G., & McComas, D. J. 2023d, Phys. Scr., 98, 105605
Livadiotis G. & McComas D. J. 2024a, NatSR, 14, 22641
Livadiotis, G., & McComas, D. J. 2024b, EPL, 146, 41003
Livadiotis G. & McComas D. J. 2025, NatSR, Submitted
Livadiotis G., McComas, D. J., Dayeh, M. A., Funsten, H. O., & Schwadron, N. A. 2011, ApJ, 734, 1
Livadiotis G., Assas, L., Dennis, B., Elaydi, S., & Kwessi, E. 2016, NRM, 29, 130
Livadiotis, G., Desai, M. I., & Wilson, L. B. III 2018, ApJ, 853, 142
Livadiotis G., Dayeh, M. A., & Zank, G. 2020, ApJ, 905, 137
Livadiotis G., Nicolaou, G., & Allegrini F. 2021, ApJ, 253, 16
Livadiotis, G., McComas, D. J., Funsten, H. O., et al. 2022, ApJS, 262, 53
Livadiotis, G., McComas, D. J., & Zirnstein, E. J. 2023, 951, 21
Livadiotis, G., Cummings, A. T., Cuesta, M. E., et al. 2024a, ApJ, 973, 6
Livadiotis, G., McComas, D. J., & Shrestha, B. L. 2024b, ApJ, 968, 66
Maksimovic, M., Pierrard, V., & Lemaire, J. F. 1997, A&A, 324, 725
Mann, G., Classen, H. T., Keppler, E., & Roelof, E. C. 2002, A&A, 391, 749
Marsch, E. 2006, LRSP, 3, 1

Maxwell, J. C. 1860, PMag, 19, 19
McComas, D. J., Allegrini, F., et al. 2009a, SSRv, 146, 11
McComas, D. J., Allegrini, F., Bochsler, P., et al. 2009b, Sci, 326, 959
McComas, D. J., Alexander, N., Angold, N. et al. 2016, *SSRv*, 204, 187
McComas, D. J., Christian, E. R., Cohen, C. M. S., et al. 2019, Natur, 576, 223
Meyer-Vernet, N., Moncuquot, M., & Hoang, S. 1995, Icar, 116, 202
Milovanov, A. V., & Zelenyi, L. M. 2000, NPGeo, 7, 211
Mitchell, J. G., de Nolfo, G. A., Hill, M.E., et al. 2020, ApJ, 902, 20
Nelson, K. P. et al. 2017, Physica A, 468, 30
Newbury, J.A., Russell, C.T., Lindsay, G.M.: 1997, Geophys. Res. Lett. 24, 1431
Nicolaou, G., & Livadiotis, G. 2016, Ap&SS, 361, 359
Nicolaou, G., & Livadiotis, G. 2019, ApJ, 884, 52
Nicolaou, G., Livadiotis, G., & Moussas, X. 2014, SoPh, 289, 1371
Nicolaou, G., Livadiotis, G., & Wicks, R. T. 2019, Entrp, 21, 997
Nicolaou, G., Livadiotis, G., Wicks, R. T., Verscharen, D., & Maruca, B. A. 2020, ApJ, 901, 26
Nicolaou, G., Livadiotis, G., & Desai, M.I. 2021, Appl. Sci., 11, 4643
Olbert, S. 1968, in Physics of the Magnetosphere, ed. R. L. Carovillano, J. F. McClay, & H. R. Radoski (New York: Springer), 641
Ogilvie, K. W., Chornay, D. J., Fritzenreiter, R. J., et al. 1995, SSRv, 71, 55
Ourabah, K. 2024, PRE, 109, 014127
Palmerio, E., Luhmann, J. G., Mays, M. L., et al. 2024, JSWSC, 14, 3
Pavlos, G. P., Malandraki, O. E., Pavlos, E. G., et al. 2016, PhysA, 464, 149
Peterson, J., Dixit, P. D., & Dill, K. A. 2013, PNAS, 110, 20380
Pierrard, V., & Lazar, M. 2010, SoPh, 267, 153
Sackur, O. 1911, *AnPhy*, 36, 958
Sarlis, N., Livadiotis, G., McComas, D.J., Cuesta, M.E., Khoo, L.Y., Cohen, C.M.S., et al., 2023, ApJ, 969, 64
Schwadron, N. A., Dayeh, M., Desai, M., et al. 2010, ApJ, 713, 1386
Schwadron, N. A., Bale, S., Bonnell, J., et al. 2020, ApJS, 246, 33
Silveira, F. E. M., Benetti, M.H., & Caldas, I.L. 2021, SoPh, 296, 113
Skellam, J. G. 1946, *J. Royal Stat. Soc. A*, 109, 296
Tathe, K., & Ghosh, S. 2024, arXiv:2407.19227v1
Tetrode, O. 1912, *AnPhy*, 38, 434
Treumann, R. A. 1997, GRL, 24, 1727
Tsallis, C. 1988, JSP, 52, 479
Tsallis, C. 2019, Entrp, 21, 696
Tsallis, C. 2023, Introduction to Nonextensive Statistical Mechanics (New York: Springer)
Vasyliũnas, V. M. 1968, JGRA, 73, 2839
Wiedenbeck, M. E., Bučík, R., Mason, G.M., et al. 2020, ApJS, 246, 42
Wilson, L. B., III, Chen, L.-J., Wang, S., et al. 2019, ApJS, 243, 8
Yoon, P. H. 2014, JGRA, 119, 7074
Yoon, P. H. 2019, Classical Kinetic Theory of Weakly Turbulent Nonlinear Plasma Processes (Cambridge: Cambridge Univ. Press)
Yoon, P. H., Ziebell, L. F., Gaelzer, R., Lin, R. P., & Wang, L. 2012, SSRv, 173, 459
Yoon, P. H., López, R. A., Salem, C. S., Bonnell, J. W., Kim, S. 2024, Entrp, 26, 310
Zank, G. P., Li, G., Florinski, V., et al. 2006, JGRA, 111, A06108
Zouganelis, I. 2008, JGR, 113, A08111